\documentclass[twocolumn,showpacs,preprintnumbers,amsmath,amssymb]{revtex4}

\usepackage{graphicx}% Include figure files
\usepackage{dcolumn}% Align table columns on decimal point
\usepackage{bm}% bold math

\begin{document}

\title{On a New Type of Information Processing \\
for Efficient Management of Complex Systems}

\author{Victor Korotkikh}
\email{v.korotkikh@cqu.edu.au}

\affiliation{School of Computing Sciences\\
Central Queensland University\\
Mackay, Queensland, 4740\\
Australia}

\author{Galina Korotkikh}

\email{g.korotkikh@cqu.edu.au}

\affiliation{School of Computing Sciences\\
Central Queensland University\\
Mackay, Queensland, 4740\\
Australia}

%\date{\today}

\begin{abstract}

It is a challenge to manage complex systems efficiently without
confronting NP-hard problems. To address the situation we suggest
to use self-organization processes of prime integer relations for
information processing. Self-organization processes of prime
integer relations define correlation structures of a complex
system and can be equivalently represented by transformations of
two-dimensional geometrical patterns determining the dynamics of
the system and revealing its structural complexity. Computational
experiments raise the possibility of an optimality condition of
complex systems presenting the structural complexity of a system
as a key to its optimization. From this perspective the
optimization of a system could be all about the control of the
structural complexity of the system to make it consistent with the
structural complexity of the problem. The experiments also
indicate that the performance of a complex system may behave as a
concave function of the structural complexity. Therefore, once the
structural complexity could be controlled as a single entity, the
optimization of a complex system would be potentially reduced to a
one-dimensional concave optimization irrespective of the number of
variables involved its description. This might open a way to a new
type of information processing for efficient management of complex
systems.

\end{abstract}


\maketitle

\begin{figure}
\includegraphics[width=.51\textwidth]{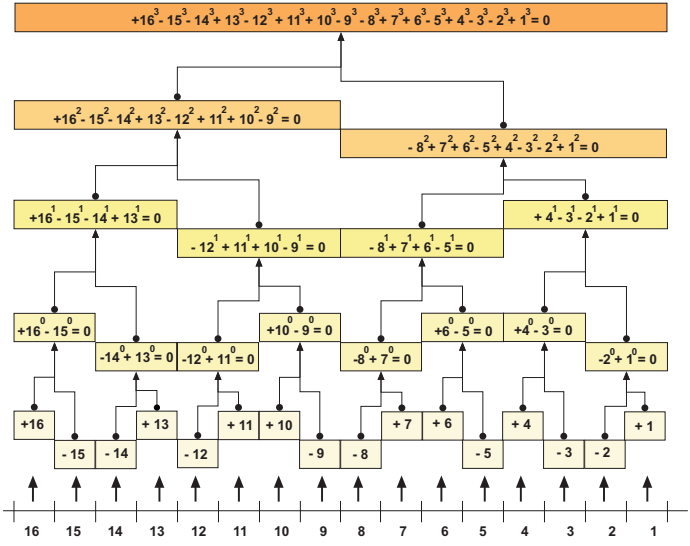}
\caption{\label{fig:one} The figure shows a hierarchical structure
of prime integer relations defined as a result of a
self-organization process starting at the zero level with integers
16, 13, 11, 10, 7, 6, 4, 1 in the "positive state" and integers 15,
14, 12, 9, 8, 5, 3, 2 in the "negative". The process is controlled
by arithmetic and cannot progress to level $5$, because arithmetic
determines that
$+16^{4}-15^{4}-14^{4}+13^{4}-12^{4}+11^{4}+10^{4}-9^{4}
-8^{4}+7^{4}+6^{4}-5^{4}+4^{4}-3^{4}-2^{4}+1^{4} \neq 0$. This
hierarchical structure can specify a correlation structure of a
system made of $16$ elementary parts ${P_{i}, i = 1,...,16}$ whose
changes ${\Delta x_{i}, i = 1,...,16}$ of the observables ${x_{i}, i
= 1,...,16}$ relative to their reference frames can be given by the
sequence $\Delta x_{1}...\Delta x_{16} =
+1-1-1+1-1+1+1-1-1+1+1-1+1-1-1+1$.}
\end{figure}

\section{Introduction}

It is a challenge to manage complex systems efficiently without
confronting NP-hard problems. To address the situation we consider
the description of complex systems in terms of self-organization
processes of prime integer relations \cite{Korotkikh_1} and suggest
to use the processes for information processing.

\section{The Hierarchical Network of Prime Integer Relations}

The description is realized through the unity of two equivalent
forms, i.e., arithmetical and geometrical. We briefly present the
forms in order to introduce the hierarchical network of prime
integer relations. More details may be found in
\cite{Korotkikh_1}, \cite{Korotkikh_2}.

\subsection{THE ARITHMETICAL FORM}

In the arithmetical form a complex system is characterized by
hierarchical correlation structures built in accordance with
self-organization processes of prime integer relations. As each of
the correlation structures is ready to exercise its own scenario
and there is no mechanism specifying which of them is going to
take place, an intrinsic uncertainty about the complex system
exists. At the same time, the information about the correlation
structures can be used to evaluate the probability of an
observable to take each of the measurement outcomes. Therefore,
the arithmetical form of the description provides the statistical
information about a complex system.

The form reveals nonlocal correlations without reference to signals
as well as the distances and local times of the parts. Thus, the
arithmetical form suggests that parts of a complex systems may be
far apart in space and time and yet remain interconnected with
instantaneous effect on each other, but no signalling. Namely, if a
correlation structure of a system is selected and some parts are
specified, then through the prime integer relations in control of
the correlation structure the other parts are immediately defined.

Through the arithmetical form of the description we become aware of
the hierarchical network, i.e., a set of mutually self-consistent
elements built by all self-organization processes of prime integer
relations. Arithmetic ensures that not even a minor change can be
made to any element of the network (Figure 1). It is helpful to
think that self-organization processes of prime integer relations
take place in the hierarchical network as they compose more and more
complex structures.

The hierarchical network of prime integer relations is a causal
structure. As parts of a system change and the prime integer
relations in control of the system cause the other parts to change
accordingly, an event takes place. Once the changes have been
realized, the event is fixed in space and time with respect to the
reference frames of the parts. For the parts the effect of the
event has not necessarily be the same, but for each part it is
appropriately determined by the prime integer relations. However,
the prime integer relations at work for a system have no causal
power to effect systems controlled by separate prime integer
relations. As a result, information about the systems is blocked
for the observers of the system.

\subsection{THE GEOMETRICAL FORM}

Specified by two parameters $\varepsilon > 0$ and $\delta > 0$ the
geometrical form arises as the self-organization processes of prime
integer relations find isomorphic realization in terms of
transformations of two-dimensional geometrical patterns
\cite{Korotkikh_1}. As a result, hierarchical structures of prime
integer relations defining the correlation structures of a complex
system become equivalently represented by hierarchical structures of
geometrical patterns determining the dynamics of the system and
revealing its complexity. The quantitative description of the system
turns out to be given by the description of the geometrical patterns
\cite{Korotkikh_1}, \cite{Korotkikh_2}.

\begin{figure}
\includegraphics[width=.45\textwidth]{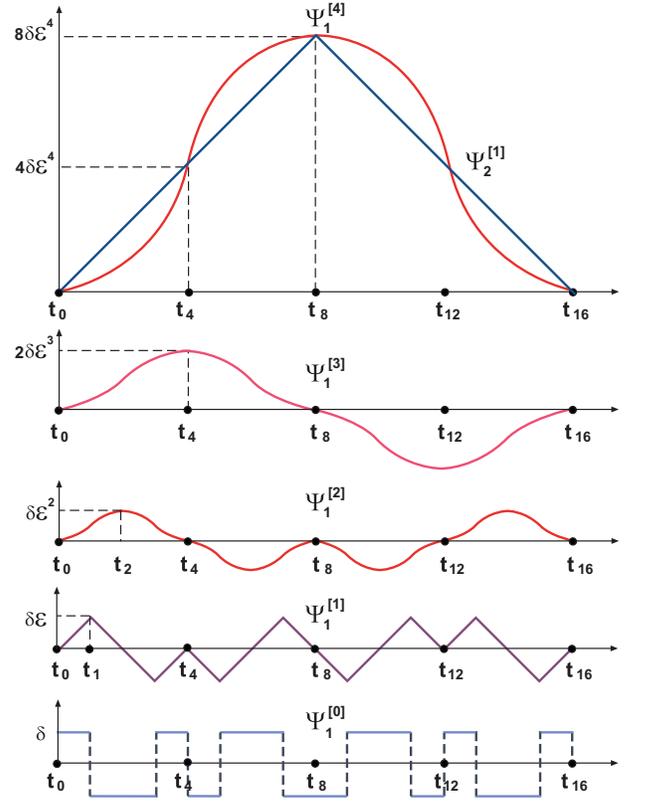}
\caption{\label{fig:two} The figure shows a hierarchical structure
of geometrical patterns, which for given $\varepsilon$ and $\delta$
is isomorphic to the hierarchical structure of prime integer
relations shown in Figure 1. Under the integration of the function
$\Psi_{1}^{[l]}, \Psi_{1}^{[l+1]}(t_{0}) = 0, l = 0,1,2,3$ the
geometrical patterns of the level $l$ form the geometrical patterns
of the higher level $l+1$. The width $W_{l}$ of a geometrical
pattern at level $l, l = 1,...,4$ equals the sum $W_{l} =
W_{l-1,left} + W_{l-1,right} = 2W_{l-1}$ of the widths of the two
geometrical patterns it is made up of, where $W_{l-1,left}$ and
$W_{l-1,right}$ are the widths of the left and right geometrical
patterns and $W_{l-1,left} = W_{l-1,right} = W_{l-1}$ as each
geometrical pattern at level $l-1$ has the same width $W_{l-1} =
2^{l-1}\varepsilon$ and $t_{i} = \varepsilon i, i = 0,1,...,16$. The
height $H_{l}$ of the geometrical pattern equals $H_{l} = S_{l-1}$,
where $S_{l-1}$ is the area of each of the geometrical patterns it
is composed of. The width $W_{l}$ and the height $H_{l}$ can specify
length scales of the level $l, l = 0,1,...,4$.}
\end{figure}

Figure 2 shows a hierarchical structure of geometrical patterns,
which for given $\varepsilon > 0$ and $\delta
> 0$ is isomorphic to the hierarchical structure of prime integer
relations depicted in Figure 1. A scale invariant property and a
renormalization group transformation come to our attention, as we
consider the connection between a geometrical pattern and the
geometrical patterns it is made of.

Although the geometrical patterns are not triangles at level $l, l
= 2,3,4$, yet the boundary curve, as the graph of the function
$\Psi_{1}^{[l]}$, is such that the area $S_{l}$ of a geometrical
pattern can be simply given, as if it were a triangle, by
$$
S_{l} = \frac{W_{l}H_{l}}{2},
$$
where $W_{l}$ and $H_{l}$ are its width and height.

The renormalization group transformation at level $4$ defines
$$
\varepsilon' = 2^{3}\varepsilon, \ \ \ \delta' =
\varepsilon^{3}\delta
$$
and uses a coarse-grained procedure replacing the geometrical
pattern made up of $8$ geometrical patterns at level $1$ by their
enlarged version with the boundary curve as the graph of the
function $\Psi_{2}^{[1]}$. Each of the geometrical patterns at level
$1$ is elementary in the sense that it is fully specified by a prime
integer relation made up of two integers without further internal
structure. The width $2\varepsilon'$ and the height
$\delta'\varepsilon'$ of the renormalized geometrical pattern at
level $4$ are given in terms of $\varepsilon'$ and $\delta'$ in the
same way as the width $2\varepsilon$ and the height
$\delta\varepsilon$ of a geometrical pattern at level $1$ are given
in terms of $\varepsilon$ and $\delta$. The two geometrical patterns
at level $4$ have the same width, height and area
$$
\int_{t_{0}}^{t_{16}} \Psi^{[4]}_{1}(t)dt = \int_{t_{0}}^
{t_{16}} \Psi^{[1]}_{2}(t)dt,
$$
but the lengths of their boundary curves are different (Figure 2).

The arithmetical and geometrical forms unite the dynamics and the
structure of a complex system as two sides of the same entity in the
preservation of the system as a whole. In particular, at one side
the dynamics of the parts is determined to produce spacetime
patterns of the parts to fit precisely into the geometrical patterns
of the complex system. Under this condition at the side the
corresponding prime integer relations can provide the relationships
between the parts for the complex system to exist. If the spacetime
patterns do not fit even slightly, then one or more of the
relationships are not in place and the complex system collapses.

To measure the complexity of a system in terms of self-organization
processes of prime integer relations a concept of structural
complexity is introduced \cite{Korotkikh_1}. Starting with the
integers at the zero level, the self-organization processes of prime
integer relations progress to different levels and thus produce a
hierarchical complexity order. In particular, the higher the level
self-organization processes progress to, the greater is the
structural complexity of a corresponding system. In our description
systems can be compared in terms of complexity by using two
equivalent forms: in structure - by the hierarchical structures of
prime integer relations and in dynamics - by the hierarchical
structures of geometrical patterns.

Remarkably, based on integers and controlled by arithmetic only
self-organization processes of prime integer relations can describe
complex systems by information not requiring further simplification.

\section{Testing a new medium for information processing}

The correlation structures of a complex system contain information
about the parts. By changing some parts the information can be
processed as the other parts change in accordance with the
correlation structures. This shows the importance of
self-organization processes of prime integer relations for
information processing. Namely, for a given problem they could be
used to build the correlation structures of a system in processing
information demonstrating the optimal performance for the problem.

As a result, we suggest the hierarchical network of prime integer
relations as a new medium for information processing and are
interested in its navigating properties. It would be important if
for any problem the performance of a system could behave as a
concave function of its structural complexity. Guided by this
property the performance global maximum for a problem could be
efficiently found. It would be also beneficial if at the global
maximum the structural complexities of the system and the problem
could be related through an optimality condition.

The optimality condition might be interpreted as follows: if through
arithmetical interdependencies emerging in the hierarchical network
between a computing system and a problem a new building block is
formed at the highest possible level, then the optimal performance
takes place. Or, in other terms, a computing system finds the
solution to a problem, once arithmetic interdependencies emerging
between the system and the problem provide a channel to obtain the
desired information.

It is worth to note that since the correlation structures of a
system are completely determined by prime integer relations, which
are equivalent to two-dimensional geometrical patterns, the entropy
of the system, measuring its information content, can be connected
with the areas of the two-dimensional patterns. Thus, in our
approach there is a general connection between entropy and area.

Computational experiments have been conducted \cite{Korotkikh_3} to
test the navigating properties. In particular, an optimization
algorithm ${\cal A}$, as a complex system, of $N$ computational
agents minimizing the average distance in the travelling salesman
problem (TSP) is developed. The agents work in parallel and start in
the same city by choosing the next city at random. Then an agent at
each step visits the next city by using one of the two strategies: a
random strategy or the greedy strategy.

In the solution of a problem with $n$ cities the state of the
agents at step $j, j = 1,...,n-1$ can be described by a binary
sequence $s_{j} = s_{1j} ...s_{Nj}$, where $s_{ij} = +1$, if agent
$i, i = 1,...,N$ uses the random strategy and $s_{ij} = -1$, if
the agent $i$ uses the greedy strategy. The dynamics of the
complex system is realized as the agents step by step choose their
strategies and can be encoded by an $N \times (n-1)$ binary
strategy matrix
$$
S = \{s_{ij}, i = 1,...,N, j = 1,...,n-1\}.
$$

We try to change the structural complexity of the algorithm ${\cal
A}$ monotonically by forcing the system to make the transition
from regular behaviour to chaos by period-doubling. To control the
system in this transition a parameter $v,  0 \leq v \leq 1$ is
introduced. It specifies a threshold point dividing the interval
of current distances passed by the agents into two parts, i.e.,
successful and unsuccessful. This information is provided for an
optimal if-then rule \cite{Korotkikh_1} that each agent uses to
choose the next strategy. The rule relies on the
Prouhet-Thue-Morse (PTM) sequence
$$
+1-1-1+1-1+1+1-1 \ . . .
$$
and has the following description:

1. if the last strategy is successful, continue with the same
strategy.

2. if the last strategy is unsuccessful, consult PTM generator
which strategy to use next.

Remarkably, for each problem $p$ tested from a class ${\cal P}$ it
has been found that the performance of the algorithm ${\cal A}$
indeed behaves as a concave function of the control parameter with
the only global maximum at a value $v^{*}(p)$. The global maximums
$\{v^{*}(p), p \in {\cal P}\}$ are of interest to probe whether
the structural complexities of the algorithm ${\cal A}$ and the
problem are related through an optimality condition. For this
purpose strategy matrices
$$
\{S(v^{*}(p)), p \in {\cal P}\}
$$
corresponding to the global maximums $\{v^{*}(p), p \in {\cal
P}\}$ are used and the structural complexities of the algorithm
${\cal A}$ and a problem $p$ are approximated as follows. The
structural complexity ${\bf C}(A(p))$ of the algorithm ${\cal A}$
is approximated by the quadratic trace
%\small
$$
C({\cal A}(p)) = \frac{1}{N^{2}}tr(V^{2}(v^{*}(p))) =
\frac{1}{N^{2}}\sum_{i=1}^{N} \lambda_{i}^{2}
$$
\normalsize of the variance-covariance matrix $V(v^{*}(p))$
obtained from the strategy matrix $S(v^{*}(p))$, where
$\lambda_{i}, i = 1,...,N$ are the eigenvalues of $V(v^{*}(p))$.

The structural complexity ${\bf C}(p)$ of the problem $p$ is
approximated by the quadratic trace
$$
C(p) = \frac{1}{n^{2}}tr(M^{2}(p)) = \frac{1}{n^{2}}\sum_{i=1}^{n}
(\lambda_{i}')^{2}
$$
of the normalized distance matrix
$$
M(p) = \{ d_{ij}/d_{max}, i,j
= 1,...,n \},
$$
where $\lambda_{i}', i = 1,...,n$ are the eigenvalues of $M(p)$,
$d_{ij}$ is the distance between cities $i$ and $j$ and $d_{max}$
is the maximum of the distances.

To reveal the optimality condition the points with the coordinates
$$
\{x = C(p), y = C({\cal A}(p)), p \in {\cal P} \}
$$
are considered. The result has been indicative of a linear
relationship between the structural complexities and thus suggests
an optimality condition of the algorithm ${\cal A}$
\cite{Korotkikh_3}:
\smallskip

{\it If the algorithm ${\cal A}$ demonstrates the optimal
performance for a problem $p$, then the structural complexity
$C({\cal A}(p))$ of the algorithm ${\cal A}$ is in the linear
relationship
$$
C({\cal A}(p)) = 0.67 C(p) + 0.33
$$
with the structural complexity $C(p)$ of the problem $p$.}
\smallskip

According to the optimality condition if the optimal performance
takes place, then in terms of the structural complexity the dynamics
of the algorithm ${\cal A}$ is in a certain relation with the
structure of the problem $p$, i.e., the distance network with the
vertices as the cities and the edges specifying the pairwise
distances.

The optimality condition is a practical tool. For a given problem
$p$ by using the distance matrix we can calculate the structural
complexity $C(p)$ of the problem $p$ and from the optimality
condition find the structural complexity
$$
C({\cal A}(p)) = 0.67 C(p) + 0.33.
$$
Then to obtain the optimal performance of the algorithm ${\cal A}$
for the problem $p$ we need to tune the control parameter for the
algorithm ${\cal A}$ to function with the structural complexity
$C({\cal A}(p))$.

\section{The order of the processes and efficient quantum algorithms}

The computational results point that by using self-organization
processes of prime integer relations it may be possible to design
classical algorithms comparable to efficient quantum algorithms.

Quantum algorithms rely on the practical use of  entanglement, whose
sensitivity challenges the development of relevant technologies. In
a TSP quantum algorithm the wave function would be evolved to
maximize through the amplitudes the probability of the shortest
routes to be measured. However, there is no general direction known
for the evolution to take in order to make the quantum algorithm
efficient. In this regard the majorization principle seems to play
an important role \cite{Latorre_1}. It provides a local navigation,
but without information about the whole landscape.

While the nature of quantum entanglement is yet to be understood
\cite{Gisin_1} in our approach the nonlocal correlations are known
from their origin in the self-organization processes of prime
integer relations. The question is whether this knowledge could be
used to provide computational resources comparable to quantum
computation.

In this paper we focus on whether such a resource could be obtained
from the nonlocal correlations although used in classical
computation, but with the order of the processes preserved. If that
could be possible, then the order of the self-organization processes
would establish a general direction for efficient computation.
Remarkably, the experiments raise the possibility that if the
evolution goes in this direction, then the performance landscape
becomes concave.

In the experiments we use the parameter to control the correlation
structures of the computing system with their consequences
observed in the routes taken by the agents. To help in
associations with the quantum case the average distance for a
value $v$ of the parameter can be written as
$$
\bar D = \frac{1}{N}(\gamma_{1,...,n-1}(v)d([1,...,n-1>) + ...
$$
$$
+ \gamma_{n-1,...,1}(v)d([n-1,...,1>))
$$
where $\gamma_{i_{1},...,i_{n-1}}(v)$ is the number of the agents
followed the route $[i_{1},...,i_{n-1}>, d([i_{1},...,i_{n-1}>)$
is its distance and the $n$ cities are labelled by $0,1,...,n-1$
with the initial city by $0$. The interpretation of the
coefficient
$$
\frac{\gamma_{i_{1},...,i_{n-1}}(v)}{N}
$$
as the probability of the route $[i_{1},...,i_{n-1}>$ suggests
that the minimization of the average distance considered in the
algorithm ${\cal A}$ can be connected with the maximization of the
probability of the shortest route considered in TSP quantum
algorithms.

Significantly, the experiments have shown that in the case of the
algorithm ${\cal A}$ the maximization of the probability turns out
to be a one-dimensional concave optimization. In its course the
computing system adapting to the problem got more complex or simpler
until the global maximum is reached and the structural complexities
of the computing system and the problem become related through the
optimality condition. More connections might arise if the wave
function could be involved in the description of the correlation
structures.

We note that the algorithm ${\cal A}$ shares a common feature with
Shor's algorithm, which also relies on the PTM sequence
\cite{Maity_1}.

\section{Conclusions}

We have suggested that self-organization processes of prime integer
relations could be used for information processing.

In particular, for a given problem self-organization processes of
prime integer relations could be used to build the correlation
structures of a system in processing information demonstrating the
optimal performance for the problem. Remarkably, the processes can
be equivalently represented by transformations of two-dimensional
geometrical patterns determining the dynamics of the system and
revealing in the information processing its structural complexity.

The information processing would be distinctive, because
self-organization processes of prime integer relations can define
the correlation structures of a system without reference to the
distances, local times and signals between the parts.

Computational experiments testing competitive advantages of the
information processing have been presented. They raise the
possibility of an optimality condition of complex systems: if the
structural complexity of a system is in a certain relationship with
the structural complexity of a problem, then the system demonstrates
the optimal performance for the problem.

The optimality condition presents the structural complexity of a
system as a key to its optimization. From its perspective the
optimization of a system could be all about the control of the
structural complexity of the system to make it consistent with the
structural complexity of the problem.

Importantly, the experiments also indicate that the performance of a
complex system may behave as a concave function of the structural
complexity. Therefore, once the structural complexity could be
controlled as a single entity, the optimization of a complex system
would be potentially reduced to a one-dimensional concave
optimization irrespective of the number of variables involved its
description. This might open a way to a new type of information
processing for efficient management of complex systems.

\bibliography{apssamp}

\end{document}